\documentclass[twocolumn,prb,twocolumn,superscriptaddress]{revtex4}
\usepackage[latin9]{inputenc}
\usepackage{color}
\usepackage{amsmath}
\usepackage{amssymb}
\usepackage{graphicx}
\usepackage{tikz}

\usepackage{mathrsfs} 
\usepackage{float}
\usepackage{bbm}

\usepackage{epsfig} 
\usepackage{verbatim}
\usepackage{array}
\usepackage{setspace}

\usepackage[unicode=true,
 bookmarks=true,bookmarksnumbered=false,bookmarksopen=false,
 breaklinks=false,pdfborder={0 0 1},backref=false,colorlinks=true]
 {hyperref}
\hypersetup{
 linkcolor=magenta, urlcolor=blue, citecolor=blue, pdfstartview={FitH}, hyperfootnotes=false, unicode=true}

\makeatletter
\@ifundefined{textcolor}{}
{%
 \definecolor{BLACK}{gray}{0}
 \definecolor{WHITE}{gray}{1}
 \definecolor{RED}{rgb}{1,0,0}
 \definecolor{GREEN}{rgb}{0,1,0}
 \definecolor{BLUE}{rgb}{0,0,1}
 \definecolor{CYAN}{cmyk}{1,0,0,0}
 \definecolor{MAGENTA}{cmyk}{0,1,0,0}
 \definecolor{YELLOW}{cmyk}{0,0,1,0}
}

\usepackage{amsfonts}\usepackage{tabularx}\usepackage{dcolumn}\usepackage{bm}\usepackage{graphicx}\usepackage{epstopdf}

\hypersetup{urlcolor=blue}

\makeatother

\newcolumntype{C}[1]{>{\centering\arraybackslash$}p{#1}<{$}}
\newcommand{\rr}{\bm{r}}
\newcommand{\Ri}{\bm{R}_i}
\newcommand{\Rj}{\bm{R}_j}
\newcommand{\Rk}{\bm{R}_k}
\newcommand{\Rl}{\bm{R}_l}

\begin{document}

\title{Suppression of charge noise using barrier control of a singlet-triplet qubit}

\author{Xu-Chen Yang}
\affiliation{Department of Physics and Materials Science, City University of Hong Kong, Tat Chee Avenue, Kowloon, Hong Kong SAR, China}
\affiliation{City University of Hong Kong Shenzhen Research Institute, Shenzhen, Guangdong 518057, China}
\author{Xin Wang}
\email{x.wang@cityu.edu.hk}
\affiliation{Department of Physics and Materials Science, City University of Hong Kong, Tat Chee Avenue, Kowloon, Hong Kong SAR, China}
\affiliation{City University of Hong Kong Shenzhen Research Institute, Shenzhen, Guangdong 518057, China}
\date{\today}

\begin{abstract}
It has been recently demonstrated that a singlet-triplet spin qubit in semiconductor double quantum dots can be controlled by changing the height of the potential barrier between the two dots (``barrier control''), which has led to a considerable reduction of charge noises as compared to the traditional tilt control method. In this paper we show, through a molecular-orbital-theoretic calculation of double quantum dots influenced by a charged impurity, that the relative charge noise for a system under the barrier control not only is smaller than that for the tilt control, but actually decreases as a function of an increasing exchange interaction. This is understood as a combined consequence of the greatly suppressed detuning noise when the two dots are symmetrically operated, as well as an enhancement of the inter-dot hopping energy of an electron when the barrier is lowered which in turn reduces the relative charge noise at large exchange interaction values. We have also studied the response of the qubit to charged impurities at different locations, and found that the improvement of barrier control is least for impurities equidistant from the two dots due to the small detuning noise they cause, but is otherwise significant along other directions.

\end{abstract}

\maketitle

\section{introduction}

The physical realization of quantum computing has attracted intensive research interest in recent years because of its potential to solve certain problems which are otherwise too difficult for a classical computer.\cite{nielsen2010quantum} Spin qubits confined in semiconductor quantum dots are among the most promising candidates for quantum computation, partially because of their demonstrated long coherence time and high control fidelities,\cite{Petta.05, Bluhm.10b, Barthel.10,Maune.12, Pla.13,Muhonen.14,Kim.14,Kawakami.16} but also due to the belief that present-day semiconductor technologies are able to extend controls on one or a few qubits to a  scaled-up array.\cite{Taylor.05}  Among the various types of spin qubits proposed theoretically\cite{Loss.98, DiVincenzo.00, Levy.02,Shi.12} and demonstrated experimentally,\cite{Petta.05, Bluhm.10b, Barthel.10,Brunner.11,Maune.12, Pla.13,Muhonen.14,Kim.14,Kawakami.16}  the singlet-triplet qubit, hosted by semiconductor Double Quantum Dot (DQD) system, stands out because it is the simplest type of spin qubits which can be controlled  solely electrostatically.\cite{Petta.05, Foletti.09,Bluhm.10c,Maune.12, Shulman.12, Wu.14, Nichol.16} 
Arbitrary single-qubit operations can be performed by combinations of $x$-axis rotations  around the Bloch sphere, which are generated by an inhomogeneous Zeeman field,\cite{Foletti.09,Bluhm.10c,Brunner.11,Petersen.13,Wu.14} and
$z$-axis rotations, accomplished by the Heisenberg exchange interaction tunable by detuning, i.e.~tilting the confinement potential (``tilt control'').\cite{Petta.05}

Two channels of noises are most destructive to the coherent operation of a singlet-triplet qubit: the nuclear, or Overhauser noise,\cite{Reilly.08, Cywinski.09a} and the charge noise.\cite{Hu.06, Nguyen.11} The nuclear noise can be substantially suppressed using dynamical Hamiltonian estimation which tracks the fluctuations in real time\cite{Shulman.14}, and can even be almost completely removed by utilization of isotropically enriched silicon in quantum-dot devices.\cite{Tyryshkin.11,Muhonen.14,Veldhorst.14} The charge noise, therefore, is now the bottleneck hindering accurate and coherent control of spin qubits.\cite{Thorgrimsson.16} The charge noise originates from unintentionally deposited impurities near the DQD system, with which electrons can hop on and off during the course of the qubit operation, creating an additional Coulomb interaction with the electrons forming the qubit. This interaction causes shifts in the energy levels of the DQD system, which subsequently leads to inaccuracies in the control field.

Very recently, it has been realized that the magnitude of the exchange interaction can alternatively be controlled by changing the height of the potential barrier in the middle of the two quantum dots (``barrier control'').\cite{Reed.16, Martins.16} While performing the barrier control, the qubit is biased to the so-called ``sweet spot'', the detuning value at which the exchange interaction is first-order insensitive to the charge noise. Therefore the charge noise can be greatly suppressed in the qubit being controlled by the potential barrier, as compared to those controlled by the traditional means of tilting. It has been experimentally demonstrated that the quality of the qubit devices increases by a factor of 5-50,\cite{Reed.16, Martins.16} suggesting the importance of the barrier control method for coherent, high-fidelity control of spin qubits. Nevertheless, the full advantage of barrier control has yet to be revealed.\cite{Zhang.17} In particular, while it is well-known that the charge noise increases with the exchange interaction for tilt control, its dependence on the exchange interaction under the barrier control is not straightforwardly clear from experimental data, which necessitates a theoretical study on the problem. 

\begin{figure}[t]
  \includegraphics[width=0.9\columnwidth]{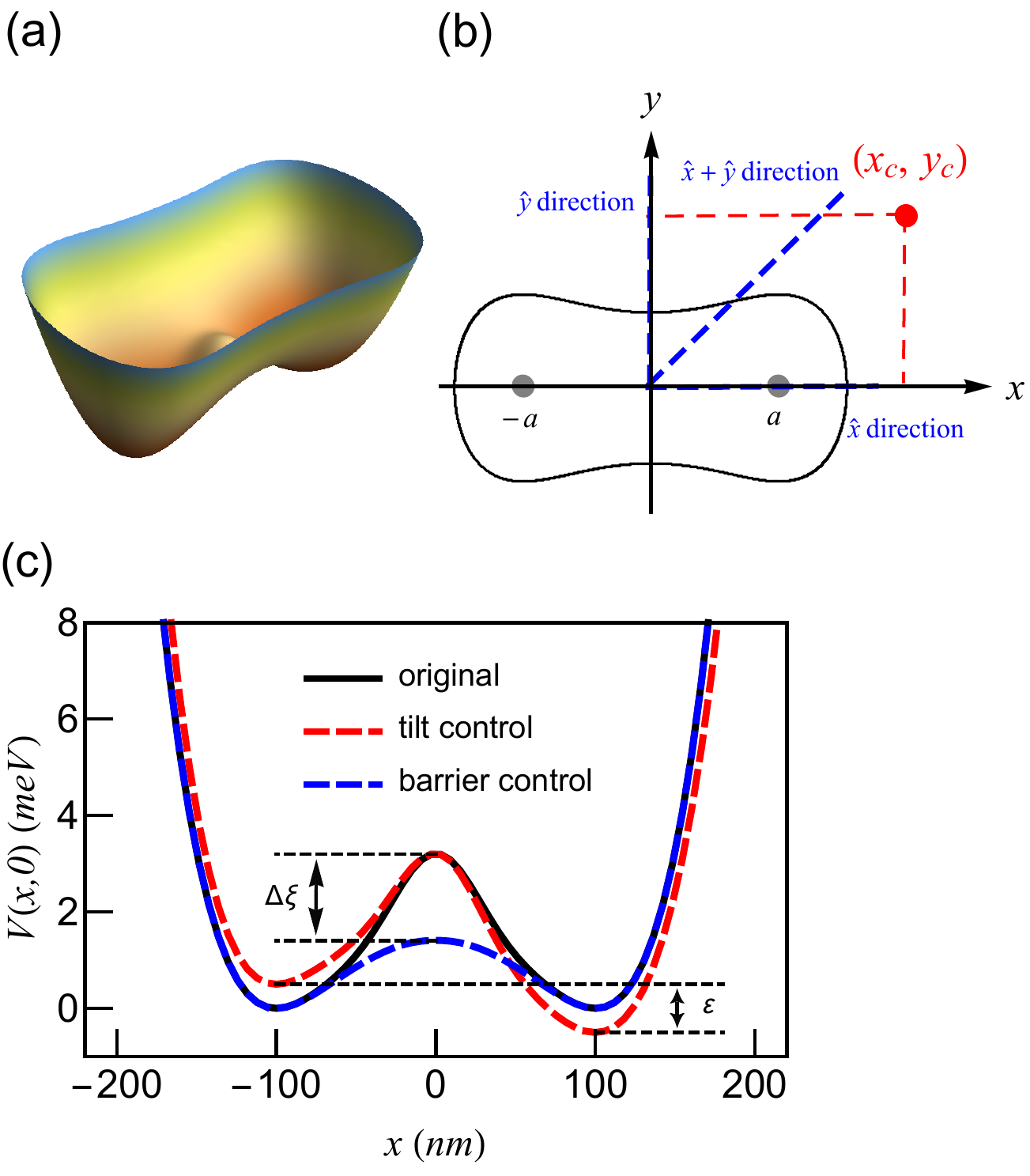}
\caption{(a) Schematic diagram of the double-well confinement potential of a DQD system, each dot occupied by one electron. (b) A charged impurity located at $\bm{R}_c=(x_c,y_c)$ shown as the red dot. Blue (thick) dashed lines show the $\hat{x}$, $\hat{y}$ and $\hat{x}+\hat{y}$ directions to be used later when discussing the effect of the impurity when it lies along one of these directions. The two wells are centered at $(\pm a, 0)$ as shown by two gray dots.
 The unit charge impurity is located on ($x_c$, $y_c$). (c)
 The confinement potential used in the calculation as described in Eq.~\eqref{eq:potential}. Black solid line shows the potential neither tilt nor barrier-controlled, which serves as the starting point as the comparison between the two control scheme. Red (asymmetric) dashed line shows the result of tilt control with detuning $\varepsilon$ with the barrier height being fixed. Blue (symmetric) dashed line shows the consequence of the barrier control while the two dots are kept leveled. Parameter: $a$ = $100$ nm, $\hbar \omega_0$ = $1.29$ meV. }
\label{fig:schematics}
\end{figure}

The molecular orbital theory has been vastly helpful in elucidating many issues arising from the development of  spin-based quantum computation.\cite{Burkard.99, Hu.00, He.05,Saraiva.07, Li.10, Yang.11b, Nielsen.12, Mehl.14,Calderon.15} In particular, calculations based on the configuration interaction method have quantified the effect of charged impurities on the energy levels of the DQD system.\cite{Nguyen.11} Further studies have shown that quantum dots each hosting multiple electrons may be controlled in a similar way to two-electron DQD,\cite{Nielsen.13} and this multi-electron singlet-triplet qubit may, in certain situations, have reduced sensitivity to charge noises thanks to the screening effect. While these works have provided important insights on our understanding of the response of a qubit to the charge noise, they have not taken into consideration the barrier control and the advantages pertaining to it. The main goal of this paper is, therefore, to provide a quantitative analysis of how the charge noise is affected depending on whether the barrier control or the tilt control method is used. We have found not only that the charge noise is much smaller for barrier control than the tilt control, but surprisingly, the relative charge noise (shift in the exchange interaction divided by its magnitude) actually \emph{decreases} with an increasing exchange interaction when the barrier control is implemented, a result that has not been appreciated in the literature.  We shall also show that this surprising fact can be understood as a result of the greatly suppressed detuning noise when the two dots are symmetrically operated, as well as the large inter-dot hopping energy of an electron when the barrier is lowered which in turn reduces the relative charge noise at large exchange interaction values. The response of the qubit to charged impurities at different locations has also been calculated.

The remainder of the paper is organized as follows. In Sec.~\ref{sec:model} we present the model and methods used in this work. We then present results in Sec.~\ref{sec:res}, including a detailed comparison of the response to charge noise under different control schemes.  We conclude in Sec.~\ref{sec:conclusion}.

\section{Model}
\label{sec:model}

Our theoretical model involves a DQD system hosting a singlet-triplet qubit, coupled to a charged impurity. The full Hamiltonian $H$ can be written as
\begin{equation}
H=H_s+H_I+H_c.
\label{eq:fullH}
\end{equation}
Here, $H_s$ is the single-electron Hamiltonian $H_s=h(\bm{r}_1)+h(\bm{r}_2)$,
\begin{equation}
h(\bm{r})=\frac{1}{2m^*}\left[\bm{p}-e\bm{A}(\bm{r})\right]^2+V(\bm{r}), 
\label{eq:sH}
\end{equation}
 where $m^*$ is the effective electron mass and $\bm{A}$ is the vector potential corresponding to the magnetic field along the $z$ direction.
$H_I$ is the Coulomb interaction between the two electrons in the DQD,
\begin{equation}
H_I=\frac{e^2}{4\pi\kappa|\bm{r}_1-\bm{r}_2|},
\label{eq:Hc}
\end{equation}
and the impurity part $H_c$ encapsulates influences on the DQD system by the impurity. 
We consider an impurity located at $\bm{R}_c=(x_c,y_c)$ in $x$-$y$ plane having charge $-e$ 
(except in Appendix~\ref{appc} where its charge is specifically noted). The impurity part of the Hamiltonian can be expressed as
\begin{equation}
H_c=\sum_{i=1}^2\frac{e^2}{4\pi\kappa|\bm{r}_i-\bm{R}_c|}.
\label{eq:im}
\end{equation}
Schematic diagrams of the confinement potential $V(\bm{r})$ and an impurity are shown in Fig.~\ref{fig:schematics}(a) and (b).

The two electrons in the DQD system form a singlet-triplet qubit. The key parameter to control the singlet-triplet qubit is the exchange interaction $J$, the energy difference between the singlet and the $S_z=0$ triplet states, which constitutes a rotation around the $z$-axis of the Bloch sphere. Traditionally the magnitude of $J$ is changed by detuning,\cite{Petta.05} namely by tilting the double well confinement potential such that one of the two dots becomes partially doubly occupied and the energy of the singlet state is changed.   Recently it has been experimentally demonstrated that $J$ can be alternatively controlled by raising and lowering the central potential barrier while keeping the two wells leveled.\cite{Reed.16, Martins.16} The barrier control method also possesses an advantage: the charge noise, which is essentially the shift of energy levels in the DQD due to nearby charged impurities, is substantially smaller compared  to that of the tilt control.

In this work we perform a microscopic calculation to compare tilt and barrier control schemes and their influence on the charge noise. To facilitate a meaningful comparison, we need a carefully designed confinement potential which can be deformed in both ways. The confinement potential is defined as
\begin{widetext}
\begin{eqnarray}
V(x,y)=
\begin{cases}
-\mu_1+\frac{m^*\omega_0^2}{2}[(x+a)^2+y^2]+\frac{4C+4\mu_1-a^2m^*\omega_0^2}{a^3}(x+a)^3+\frac{-6C-6\mu_1+a^2m^*\omega_0^2}{2a^4}(x+a)^4+G(x,y), &x\ge0\cr -\mu_2+\frac{m^*\omega_0^2}{2}[(x-a)^2+y^2]-\frac{4C+4\mu_2-a^2m^*\omega_0^2}{a^3}(x-a)^3+\frac{-6C-6\mu_2+a^2m^*\omega_0^2}{2a^4}(x-a)^4+G(x,y), &x<0\end{cases}
\label{eq:potential}
\end{eqnarray}
\end{widetext}
where $C=a^2m^*\omega_0/12$ is the height of the central potential barrier (regardless of the values of $\mu_1$ and $\mu_2$), which can be changed by $\xi$ through the Gaussian function $G(x,y)=\xi\exp\left[-8(x^2+y^2)/a^2\right]$. Alternatively the tilt control can be achieved by making 
 $\mu_{1,2}=\pm\varepsilon/2$ which is the energy in the bottom of the two potential wells. The two wells are centered at $\bm{R}_1=(-a,0)$ and $\bm{R}_2=(a,0)$ respectively, and are well approximated by the familiar harmonic oscillator potential
\begin{equation}
V(x,y)|_{(x,y)\rightarrow \bm{R}_{1/2}}\approx\frac{m^*\omega_0^2}{2}[(x\pm a)^2+y^2]-\mu_{1/2},
\label{eq:appr}
\end{equation}
where $\omega_0$ is the confinement energy which characterizes the size of the dots. Appendix~\ref{appx:Potential} provides more details on the confinement potential. In Fig.~\ref{fig:schematics}(c) we show the intersection at $y=0$ of the potential $V(x,y)$, which at the same time indicates the results of both the tilt and barrier control. Starting from the black solid line, we can increase the exchange interaction either by tilting the double well which makes the energies of the two wells different by $\varepsilon$ (``tilt control''), or lowering the central potential barrier via reducing $\xi$ (``barrier control'').

\section{Results}
\label{sec:res}

We use the molecular orbital method to characterize the electron wave functions and the energy spectrum. We apply the Hund-Mulliken approximation, in which only the ground states of a harmonic oscillator is considered:
\begin{equation}
\begin{aligned}
\phi_i(\bm{r})=\frac{1}{a_B^{}\sqrt{\pi}}\exp\left[{-\frac{1}{2a_B^2}\left|\bm{r}-\bm{R}_i\right|^2}\right],
\end{aligned}
\label{eq:ground}
\end{equation}
where $a_B\equiv\sqrt{\hbar/(m^*\omega_0)}$ is Fock-Darwin radius, and $i=1,2$ indicates the two dots respectively. The Fock-Darwin states in Eq.~\eqref{eq:ground} are then orthogonalized to give approximated single-electron wave functions in the DQD system:
\begin{equation}
\begin{aligned}
\left\{\psi_1,\text{ }\psi_2\right\}^\text{T}=\mathcal{O}^{-1/2}\left\{\phi_1,\text{ }\phi_2\right\}^\text{T},
\end{aligned}
\label{eq:trans}
\end{equation}
where $\mathcal{O}$ is the overlap matrix defined as $\mathcal{O}_{l,l^\prime}\equiv\langle\phi_l|\phi_{l^\prime}\rangle$. $\mathcal{O}^{-1/2}$ can be found, for example, following methods presented in Refs.~\onlinecite{Li.10} or \onlinecite{Yang.17}.

Without any impurity, the partial Hamiltonian of system $H_s+H_I$ [cf. Eq.~\eqref{eq:fullH}] can be written in a matrix form under the basis $\{\left|0,\uparrow\downarrow\rangle\right., \left|\downarrow\uparrow\rangle\right., \left|\uparrow\downarrow\rangle\right., \left|\uparrow\downarrow,0\rangle\right\}$, i.e. $\left|0,\uparrow\downarrow\rangle\right.=c^\dagger_{2\uparrow}c^\dagger_{2\downarrow}\left|\text{vac}\right\rangle$, $\left|\downarrow\uparrow\rangle\right.=c^\dagger_{2\uparrow}c^\dagger_{1\downarrow}\left|\text{vac}\right\rangle$, $\left|\uparrow\downarrow\rangle\right.=c^\dagger_{1\uparrow}c^\dagger_{2\downarrow}\left|\text{vac}\right\rangle$, and $\left|\uparrow\downarrow,0\rangle\right.=c^\dagger_{1\uparrow}c^\dagger_{1\downarrow}\left|\text{vac}\right\rangle$, where $c_{i\sigma}^\dagger$ creates an electron with spin $\sigma$ on the $i{\rm th}$ dot, and $\left|\text{vac}\right\rangle$ is the vacuum state. The matrix form of the Hamiltonian can then be written as\cite{Burkard.99,Yang.11,Wang.11}
\begin{equation}
\begin{split}
&H_s+H_I=\\
&\left(\begin{array}{cccc}
U_2-2\mu_2&-t&-t&0\\
-t&U_{12}-\mu_1-\mu_2&0&-t\\
-t&0&U_{12}-\mu_1-\mu_2&-t\\
0&-t&-t&U_1-2\mu_1\\
\end{array}
\right),
\label{eq:aHam1}
\end{split}
\end{equation}
where $U_{1,2}$ are on-site Coulomb interactions, $U_{12}$ is the inter-site Coulomb interaction, and $t$ is the hopping between the two quantum dots. 

\begin{figure}[t]
  \includegraphics[width=0.85\columnwidth]{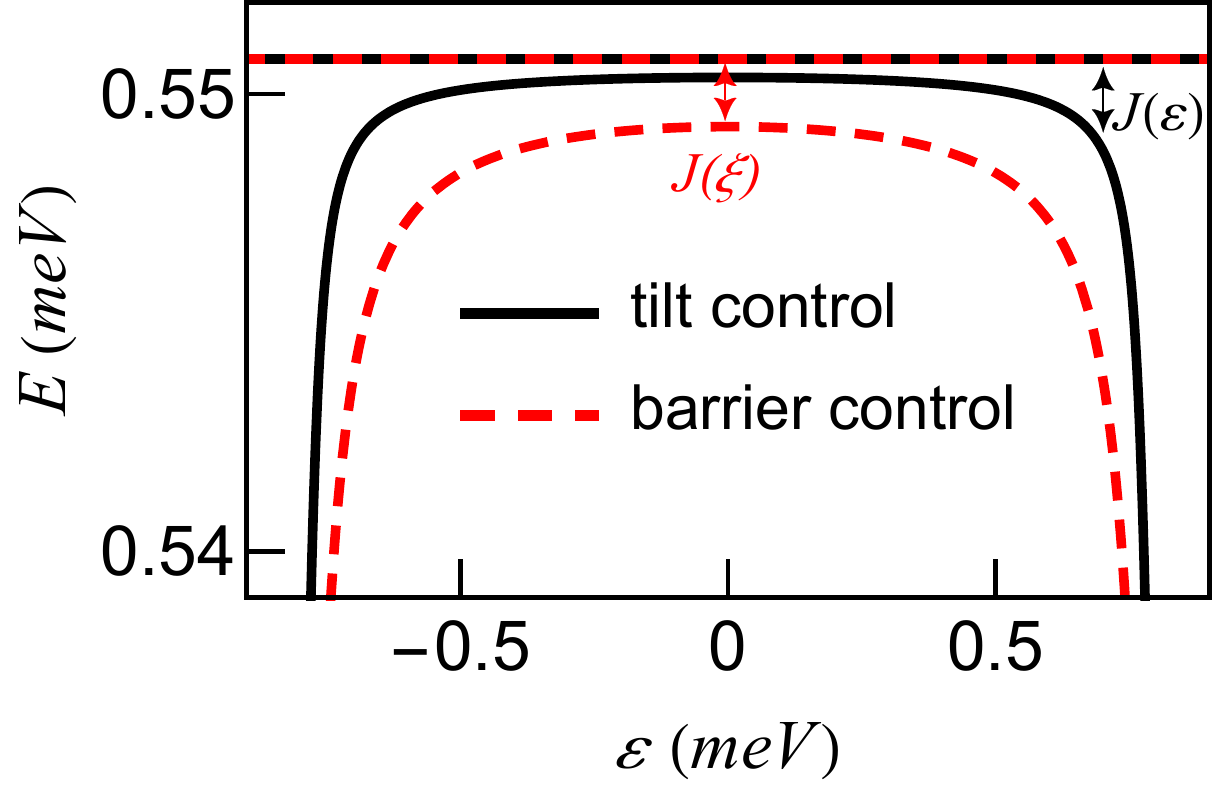}
\caption{Calculated energy spectra of the DQD system. Only the lowest two energy levels are shown. Black solid lines: energy spectrum with the middle barrier fixed by $\xi=1.3 \text{meV}$. In this case, the exchange interaction $J(\varepsilon)$ is varied by changing the detuning $\varepsilon$. Red (gray) dashed lines show the result of the barrier control of changing $\xi$ to 1 meV. (only the exchange interaction at $\varepsilon=0$ is used in the quantum computation). Parameters: $a$ = $100$ nm, $\hbar \omega_0$ = $100$ $\mu$eV. }
\label{fig:2}
\end{figure}

Matrix elements of the Hamiltonian can be evaluated by taking inner products of the relevant two-electron wave functions, which are essentially Slater determinants of orthogonalized wave functions in Eq.~\eqref{eq:trans} with appropriate spins, corresponding to the  creation operators involved in the 
second-quantized wave functions shown above. We then calculate the energy spectrum of the system by diagonalizing the Hamiltonian matrix. It is important to note that we are considering the qubit maintained at stationary to avoid complications to the calculation of eigenenergies and exchange interaction. There are prolific literature concerning the reduction of noises during the dynamical operation of qubits, including for example shortcuts to adiabaticity\cite{Yokoshi.09, Chen.10, Ban.12, Ban.14} and composite pulses.\cite{Wang.12, Kestner.13, Wang.14, Wang.15} Our results on charge noises should be regarded as inputs to those well-established methods in actual implementation of qubits.

The lowest two energy levels of the DQD system are shown in Fig.~\ref{fig:2}. The solid lines show the case of the tilt control, in which we hold $\xi$ at a constant value (1.3 meV), sufficiently large to allow enough room for enlarging $J$ in subsequent studies. In the tilt control, the exchange interaction used in the qubit manipulation is the energy difference between the two levels at various different detuning values $\varepsilon$, $J(\varepsilon)$, which is small at $\varepsilon=0$ and increases substantially as $\varepsilon$ is tuned in both positive and negative directions.  When the barrier control is used instead of detuning, the only point of interest is $\varepsilon=0$, and the distance in energy between the two levels are enlarged by decreasing $\xi$ and thereby increasing $\gamma$, shown as the dashed lines in  Fig.~\ref{fig:2}.

\begin{figure}
  \includegraphics[width=0.7\columnwidth]{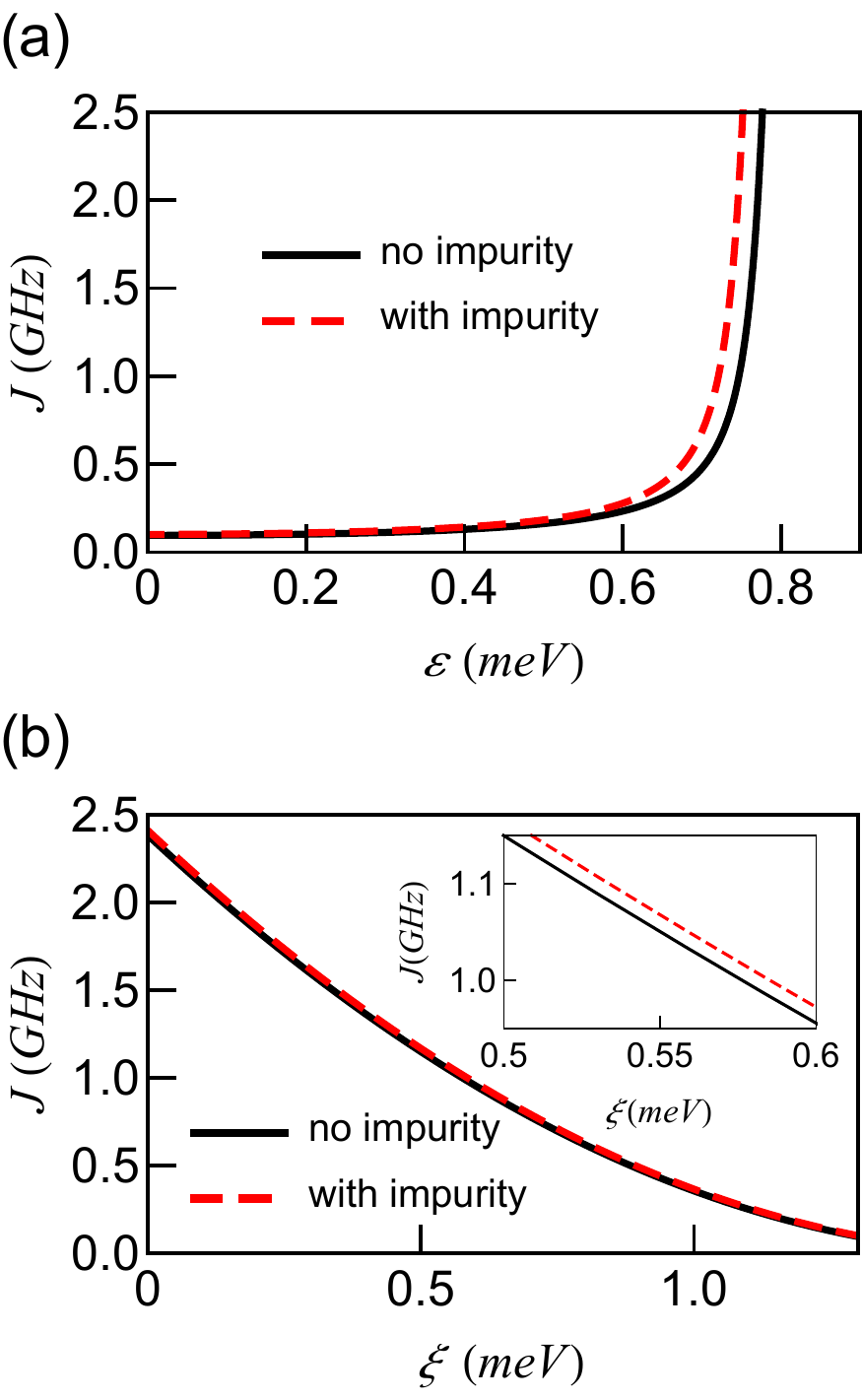}
  \caption{(a) The exchange interaction $J(\varepsilon)$ under the tilt control while the barrier is fixed by $\xi=1.3$ meV. Black solid line shows the exchange interaction without any impurity, while the red (gray) dashed line shows the case with an impurity located at $\bm{R}_c= (-6a, 6a)$. (b) The exchange interaction $J(\xi)$ under the barrier control without detuning. Black solid line: exchange interaction without any impurity; red (gray) dashed line: the case with the impurity located at the same location as (a). Inset: a zoom-in of the range 0.5 meV $\le\xi\le$ 0.6 meV.  Parameters: $a$ = $100$ nm, $\hbar \omega_0$ = $100$ $\mu$eV.}
\label{fig:3}
\end{figure}

The existence of a charged impurity adds additional terms to Eq.~\eqref{eq:aHam1} due to the Coulomb interaction between the impurity and the quantum-dot electrons. The matrix form of the full Hamiltonian $H$ is therefore
\begin{widetext}
\begin{equation}
H=\left(\begin{array}{cccc}
U_2-2\mu_2+2{Z_t}_2&-t+{Z_t}_{12}&-t+{Z_t}_{12}&0\\
-t+{Z_t}_{12}&U_{12}-\mu_1-\mu_2+{Z_t}_1+{Z_t}_2&0&-t+{Z_t}_{12}\\
-t+{Z_t}_{12}&0&U_{12}-\mu_1-\mu_2+{Z_t}_1+{Z_t}_2&-t+{Z_t}_{12}\\
0&-t+{Z_t}_{12}&-t+{Z_t}_{12}&U_1-2\mu_1+2{Z_t}_1\\
\end{array}
\right),
\label{eq:aHam2}
\end{equation}
\end{widetext}
where the $Z_t$ terms denote the corrections due to the impurity on the Hamiltonian matrix.
It is worth noting that all terms in Eq.~\eqref{eq:aHam2} can be calculated analytically thanks to the polynomial/Gaussian form of the confinement potential and the Fock-Darwin states. Explicit forms of relevant Coulomb integrals are presented in Appendix~\ref{appx:eleHam}.
The exchange interaction under the influence of such an impurity can then be evaluated again by diagonalizing Eq.~\eqref{eq:aHam2} and take the energy difference between the ground state and the first excited state. Results comparing the exchange interaction values with and without an impurity are shown in Fig.~\ref{fig:3}.  We note that for all results shown in the main text, the impurity is considered to be reasonably far away from the quantum dot. In fact, our main conclusion holds true even if the impurity is close to the quantum dot, and we show a representative result in Appendix~\ref{appc}.

\begin{figure}
  \includegraphics[width=0.75\columnwidth]{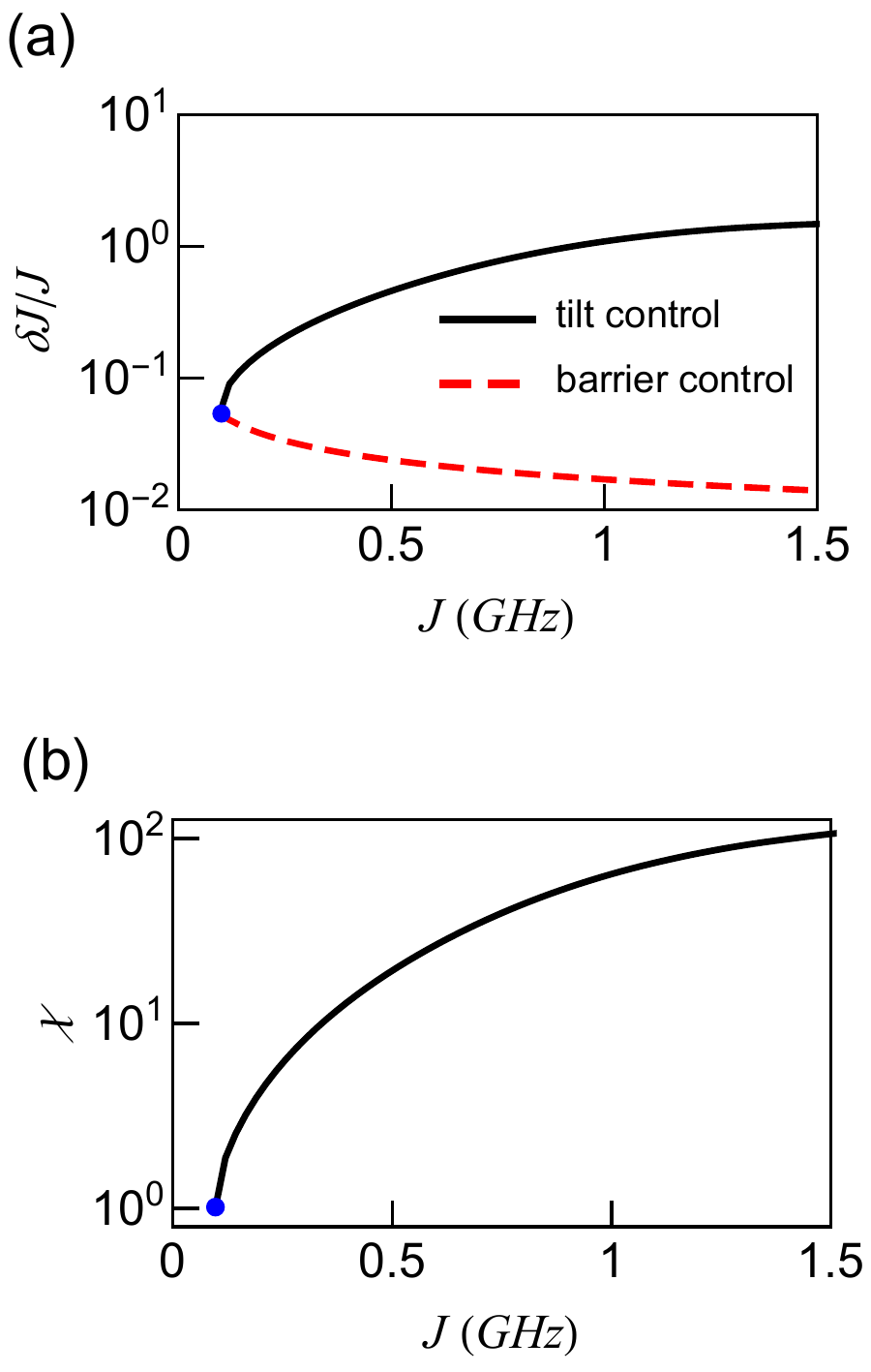}
  \caption{(a) The relative charge noise $\delta J/J$ v.s.~$J$. Black solid line: the result for the tilt control case. Red (gray) dashed line: the result under barrier control. Comparison of the different exchange interaction $\delta J/J$ caused by impurity with tilt and barrier control. Note the log scale of the $y$-axis. (b) The improvement ratio $\chi$, defined as the charge noise $\delta J$ for the tilt control divided by the value for barrier control. The impurity is positioned at $\bm{R}_c= (-6a, 6a)$.  Parameters: $a$ = $100$ nm, $\hbar \omega_0$ = $100$ $\mu$eV.}
\label{fig:4}
\end{figure}

\begin{figure}
  \includegraphics[width=0.8\columnwidth]{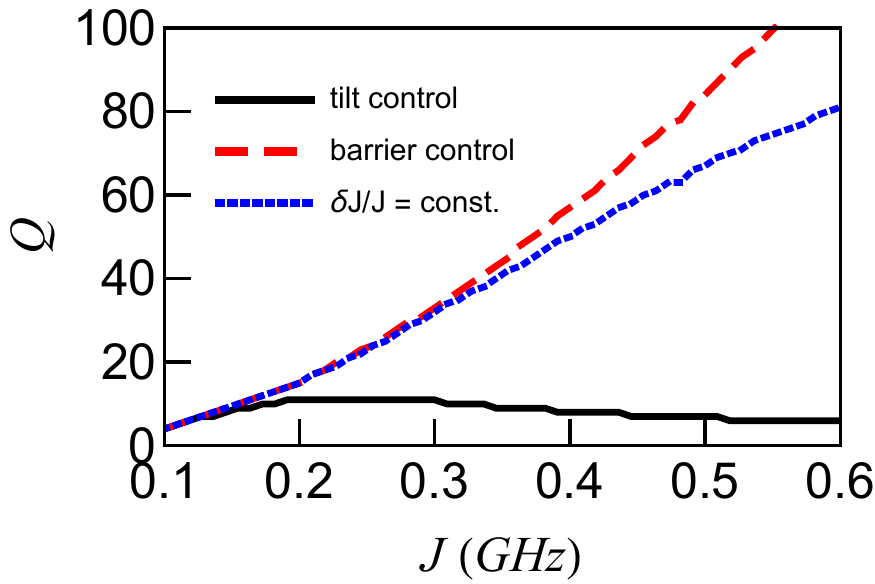}
  \caption{The quality factor $Q$ v.s.~exchange interaction $J$. Black solid line: the result for the tilt control case. Red (gray) dashed line: the result under barrier control. Blue (gray) dotted line: the result of $Q$ when the relative charge noise is assumed to be independent of the exchange interaction, i.e. $\delta J/J=\rm{const.}$
  The impurity is positioned at $\bm{R}_c= (-6a, 6a)$.  Parameters: $a$ = $100$ nm, $\hbar \omega_0$ = $100$ $\mu$eV.}
\label{fig:5}
\end{figure}

\begin{figure}
  \includegraphics[width=0.75\columnwidth]{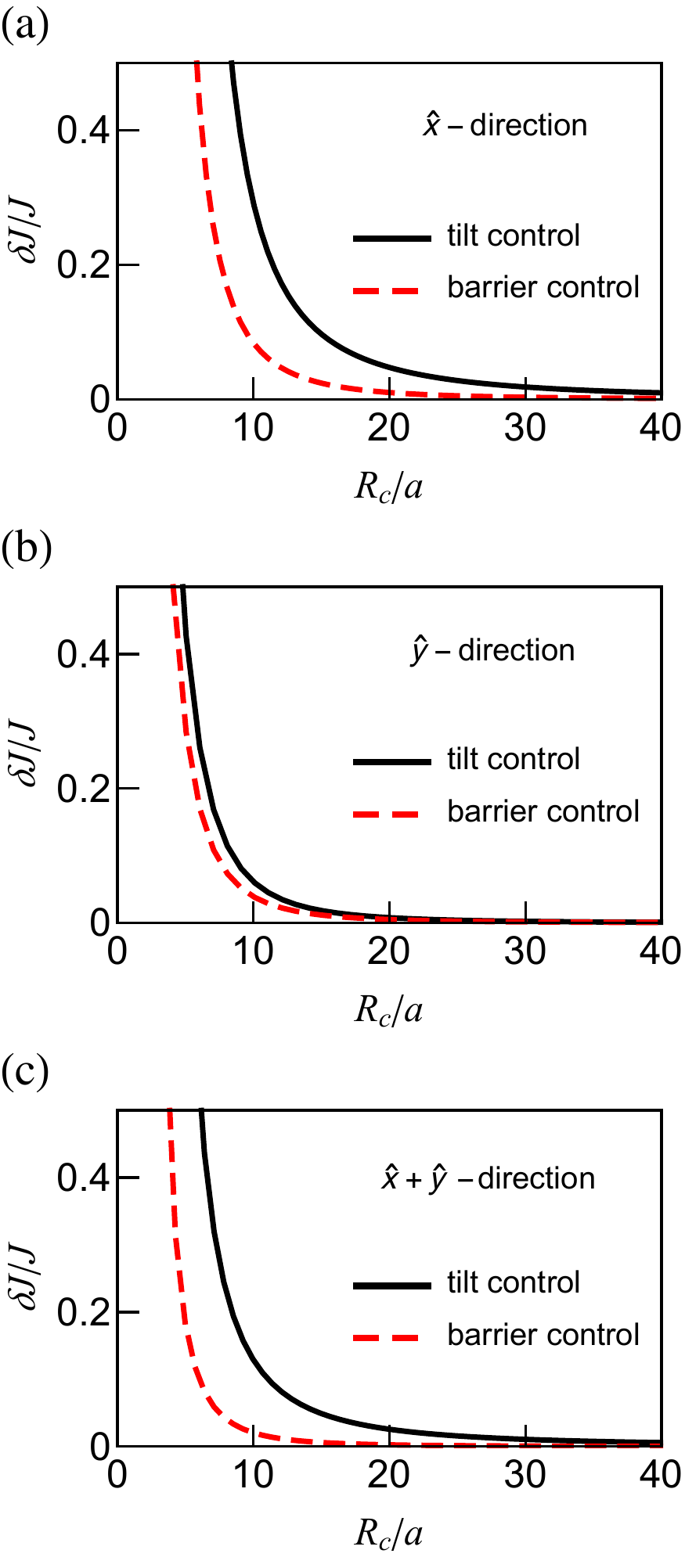}
  \caption{The relative charge noise $\delta J/J$ v.s.~the distance between the impurity and the center of the DQD system, $R_c$ = $|\bm{R}_c|$. The impurity is positioned along (a) the $\hat{x}$ direction, (b)  the $\hat{y}$ direction, and (c) the $\hat{x}+\hat{y}$ direction [cf.~the blue (gray) dashed lines of Fig.~\ref{fig:schematics}(b)]. The two control schemes are compared at $J=242$ MHz. Parameters: $a$ = $100$ nm, $\hbar \omega_0$ = $100$ $\mu$eV.}
\label{fig:6}
\end{figure}

Fig.~\ref{fig:3}(a) shows the exchange interaction under the tilt control, $J(\varepsilon)$, with the barrier fixed at $\xi=1.3$ meV. As expected, $J$ increases as the system is detuned. For small detuning $J$ does not increase much, but for large detuning ($\varepsilon>0.6$ meV) $J$ increase exponentially. In presence of an impurity, the exchange interaction is increased overall, but the increase is more pronounced for large detuning than for the small one. For large detuning ($\varepsilon>0.6$ meV) the change of $J$, which we denote as $\delta J$, is more than 10\%--30\% of $J$. This is consistent with the qualitative picture that charge noise generally increases with larger exchange interaction under tilt control.

In Fig.~\ref{fig:3}(b) we present our results on the exchange interaction under barrier control. For large $\xi$, $J$ is small, and $J$ increases as $\xi$ is decreased, corresponding to a suppression of the central potential barrier. When an impurity is present, the shift in the exchange interaction is very small (less than 1\%). This can be seen more clearly in the inset of Fig.~\ref{fig:3}(b).

To further understand how the system responds to the impurity under two different control schemes, we compare the relative charge noise $\delta J/J$ when the exchange interaction $J$ has been tuned to the same value using the two methods. The results are shown in Fig.~\ref{fig:4}(a). The comparison starts with $J=0.1$ GHz shown as the blue (gray) dot in Fig.~\ref{fig:4}(a). The confinement potential is given by  $\xi=1.3$ meV and $\varepsilon=0$, which has been shown in Fig.~\ref{fig:schematics}(c) as the solid line. From here, the exchange interaction can either be increased by tilting (increasing $\varepsilon$ but keeping $\xi=1.3$ meV), shown as the solid line in Fig.~\ref{fig:4}(a), or by lowering the potential barrier (reducing $\xi$ but keeping $\varepsilon=0$), shown as the dashed line in the same panel. For tilt control, the relative charge noise increases with the exchange interaction as expected. It is however remarkable that the relative charge noise actually decreases while the exchange interaction is increased, if the system is under barrier control. This can be understood using the following argument. The effective exchange interaction can be written, in terms of the Hubbard parameters of the Hamiltonian,\cite{Li.12,Yang.17} as
\begin{equation}
J\approx\frac{2t^2}{\Delta U+\varepsilon}+\frac{2t^2}{\Delta U-\varepsilon},
\end{equation}
where $\Delta U=U_1-U_{12}$ and note that $U_1=U_2$ for our setup of the model. 
The relative charge noise can then be expressed as
\begin{equation}
\begin{aligned}
\frac{\delta J}{J}&=\frac{1}{J}\frac{\partial J}{\partial t}\delta t+\frac{1}{J}\frac{\partial J}{\partial\varepsilon}\delta\varepsilon\\
&=\frac{2}{t}\delta t+\frac{2\varepsilon}{\Delta U^2-\varepsilon^2}\delta\varepsilon
\label{eq:dJfunct}
\end{aligned}
\end{equation} 
In the tilt control, $t$ is roughly fixed, and $\delta J/J$ clearly increases when $\varepsilon$ is increased from zero. When the DQD system is under barrier control, $\varepsilon=0$ and only the first term on the right hand side of Eq.~\eqref{eq:dJfunct} remains. In order to increase the exchange interaction one must lower the central potential barrier which consequently enlarges $t$. As a result, the charge noise is greatly suppressed. 

To reveal the advantage of the barrier control scheme we define an improvement factor $\chi$ as the relative charge noise $\delta J/J$ for the tilt control divided by the value for the barrier control scheme. $\chi$ as a function of $J$ is shown in Fig.~\ref{fig:4}(b). For the parameters of the blue (gray) dot in Fig.~\ref{fig:4}(a) (starting point for the comparison), $\chi=1$. As $J$ is increased, $\chi$ is also enhanced, indicating that the barrier control has outperformed the tilt control method by suppressing the charge noise. In a typical range of $J$ between tens and a few hundreds of MHz, $\chi$ can increase up to 10 or above, suggesting an order of magnitude reduction of the charge noise, which is consistent with the experimental observation.\cite{Reed.16, Martins.16} Further increasing $J$ beyond 1 GHz may lead to almost two orders of magnitude reduction in the charge noise, although operating the qubit at that high frequency may not be practical.

In practical experiments the reduction of charge noise is most straightforwardly uncovered by the quality factor $Q$, defined as the number of full Rabi oscillations before the amplitude decays to $1/e$ of the initial value.\cite{Reed.16, Martins.16} We therefore plot the $Q$ factor corresponding to relevant cases in Fig.~\ref{fig:5}. These results are obtained using the numerically extracted $\delta J/J$ and the quasi-static noise model discussed in Ref.~\onlinecite{Barnes.16}. It is obvious from the figure that the $Q$ factor for the tilt control is roughly a constant below 10 for a range of $J$ between 150 MHz and 300 MHz, and slightly decreases if $J$ is further increased. On the other hand, the $Q$ factor rapidly increases if the device is under barrier control. Both results are in agreement with recent experimental data.\cite{Martins.16} While we may easily draw a conclusion that the charge noise is indeed smaller for the barrier control than the tilt control, the fact that the $Q$ factor is higher for barrier control does not necessarily imply a relative charge noise which decreases with a increasing $J$. This is because $Q$ may increase for two reasons: the increase may be a result of a decreasing charge noise under which the Rabi oscillation takes more time to decay, but it could also originate from the fact that a larger $J$ implies more Rabi oscillations within the same amplitude envelope.\cite{Reed.16, Martins.16} To have a better understanding of the problem we plot the result of $Q$ as a consequence of a putative charge noise model, $\delta J/J=\rm{const.}$,\cite{Shulman.12,Dial.13} shown as the blue (gray) dotted line in Fig.~\ref{fig:5}. The constant value is again taken from the parameters of the blue (gray) dot in Fig.~\ref{fig:4}(a). The roughly linear increase of the dotted line in Fig.~\ref{fig:5} is solely due to the enhancement of the exchange interaction, and the result for the barrier control (dashed line) is above the dotted line, which clearly indicates that the relative charge noise decreases with an increasing $J$ when barrier control is implemented. 

In the results shown above we have fixed the impurity at one location, but we have verified that our main conclusions remain even when the impurity is moved around the DQD system. We show selective results in Fig.~\ref{fig:6} where we consider the relative charge noise while an impurity is moved away from the DQD along three different directions. As expected, the farther the impurity is from the center of the DQD (characterized by $R_c$), the lower the charge noise is. Moreover, in all cases that we have considered, the barrier control shows advantage over the tilt control, consistent with the results shown above. It is interesting to note from Fig.~\ref{fig:6}(b) that when the impurity is located along the $y$ axis (equidistant from the two dots), the difference in effects between the tilt and barrier control, albeit still being considerable, is rather small. This is due to the fact that the impurity causes roughly equal shifts of the energy of two potential wells even when they are detuned, resulting in a small $\delta\varepsilon$ in Eq.~\eqref{eq:dJfunct}, and leaves the first term which is only relevant to the hopping across the central potential barrier to give the main contribution.

\section{Conclusions}
\label{sec:conclusion}

In this paper, we have performed a microscopic calculation of a double quantum dot system which hosts a singlet-triplet qubit. We have focused on the effect of a charged impurity near the quantum dots, namely the charge noise, and how it behaves under two different control schemes. Traditionally, the exchange interaction is controlled by tilting the two potential wells, called the tilt control method. In recent experiments, it has been realized that the exchange interaction can alternatively be increased or decreased by lowering or raising the central potential barrier without detuning the two wells, termed as the barrier control method. It has been further observed that the barrier control method bears a particular advantage that the charge noise is substantially suppressed.\cite{Reed.16, Martins.16} From the microscopic theoretic calculations, we have provided quantitative evaluation of the extent to which the charge noise has been suppressed for qubits controlled via the barrier method as compared to the tilt control one. We have found that not only the relative charge noise is smaller for barrier control as compared to the tilt control, it in fact decreases when the exchange interaction is enlarged under the barrier control, converse to the tilt control for which the relative charge noise increases with increasing exchange interaction. For typical exchange interactions around 500 MHz, the charge noise is reduced by about an order of magnitude, and this improvement can be further increased to two orders of magnitude should the exchange interaction can be tuned beyond 1 GHz. The improvement is significant for impurities lying in most orientations with respect to the DQD system except when the impurity is equidistant from the two dots (along the $y$-axis in this work), in which case the advantage of using barrier control method is less pronounced because the impurity would cause comparable energy shifts in the two quantum wells, making the contribution from the detuning error relatively small.  Our theoretical assessment of the problem not only reaffirms the experimental observation that barrier control reduces the charge noise, it has also led to new insight that the relative charge noise actually reduces as the exchange interaction is increased, a fact that has not been sufficiently appreciated in the literature. Our results therefore constitutes an important step forward in the understanding of decoherence of spin qubits, which will eventually help in the physical realization of a scalable, fault-tolerant quantum computer.

This work is supported by the 
Research Grants Council of the Hong Kong Special Administrative Region, China (No. CityU 21300116) and the National Natural Science Foundation of China (No. 11604277).

\onecolumngrid

\appendix
\setcounter{equation}{0}

\section{The confinement potential}\label{appx:Potential}
In this section we explain the motivation behind the design of the confinement potential. First, we assume that the two dots lie along the $x$-axis and the $y$-dependence of the potential is simply a parabola centered at $y=0$, i.e.
\begin{equation}
V(x,y)=V_x(x)+\frac{1}{2}m^*\omega_0^2y^2.
\end{equation}
To facilitate the Hund-Mulliken calculation we assume that $V_x$ is a polynomial of $x$, $V_x(x)=\sum_{i=0}^{n}{b_ix^i}$.

In order to determine the parameters in the polynomial, we take into account the following considerations: The two dots must center at $x=\pm a$ respectively regardless of how they are detuned; at each of the minima the potential should resemble a parabola with minimum energy $-\mu_1$ and $-\mu_2$ respectively; and since we focus on barrier control in this work, the central barrier should be smooth and its height be fixed to certain known number $C$. These requirements are summarized as:
\begin{equation*}
\begin{array}{C{3em}|C{3em}|C{3em}|C{3em}|C{4em}|C{4.5em}}
\hline\hline
V_x(0)&V_x(-a)&V_x(a)&V_x'(0)&V_x'(\pm a)&V_x''(\pm a)\\
\hline
C&-\mu_1&-\mu_2&0&0&m^*\omega_0^2\\
\hline\hline
\end{array}
\label{eq:st2}
\end{equation*}

We have found that the simplest way to satisfy the requirement above is to define $V_x$ as two fourth order polynomials in $x$ for $x<0$ and $x\ge0$ separately and let them connect smoothly at $x=0$. The two additional orders beyond quadratic ensure that the two curves meet at $V_x=C$ and their first derivatives are continuous at $x=0$. Straightforward calculations of the parameters gives Eq.~\eqref{eq:potential} in the main text.

It is worth remarking that $C$ is not arbitrary. The second order derivative of $V_x(x=0)$, albeit being discontinuous, must not be positive for the barrier to exist. This implies that $V_x^{\prime\prime}(x,0)=(a^2m^*\omega_0^2-12\mu_{1,2}-12C)/a^2\leq0$. We keep $\mu_1=\mu_2=0$ while doing barrier control, and we have chosen $C=a^2m^*\omega_0^2/12$ in this work. 

In order to simulate the barrier control, we add a Gaussian function $G_0(x)$ to the confinement potential, $G_0(x)=\xi\cdot\exp\left[-(x-x_0)^2/2\sigma^2\right]$. We take the standard deviation $\sigma=a/4$ so that its influence will be confined to the small neighborhood around $x=0$ and will not affect the nearly quadratic shape at the bottom of the two wells. The expression of this part is therefore, 
\begin{equation}
G(x,y)=\xi\cdot\exp\left[-\frac{8(x^2+y^2)}{a^2}\right],
\label{pot:gau}
\end{equation}
as shown in Eq.~\eqref{eq:potential}.

\section{Useful results for calculating the matrix elements of the Hamiltonian}\label{appx:eleHam}

In calculating the matrix elements of Hamiltonian [Eqs.~\eqref{eq:aHam1} and \eqref{eq:aHam2}], the following integrals with analytical results are useful. For the Coulomb interaction between two electrons in the DQD system,\cite{Calderon.15}
\begin{equation}
\begin{aligned}
&\iint\phi_i(\rr_1)^*\phi_j(\rr_2)^*\frac{e^2}{4\pi\kappa|\rr_1-\rr_2|}\phi_k(\rr_1)\phi_l(\rr_2)d\rr_1d\rr_2\\
=&\frac{e^2}{4\sqrt{2\pi}\kappa a_B}\exp\left(-\frac{1}{4a^2_B}|\bm{R}_i-\bm{R}_j|^2-\frac{1}{4a^2_B}|\bm{R}_k-\bm{R}_l|^2-\frac{1}{16a^2_B}|\Ri+\Rj-\Rk-\Rl|^2\right)\\
&\times\text{I}_0\left(\frac{1}{16a^2_B}|\Ri+\Rj-\Rk-\Rl|^2\right).
\end{aligned}
\label{eq:GF2Delei}
\end{equation}
For the interaction between the impurity (having charge $-e$ and located at $\bm{R}_C$) and the electrons in the quantum dots.
\begin{equation}
\begin{aligned}
\int{\phi_i(\rr)^*\frac{e^2}{4\pi\kappa|\rr-\bm{R}_C|}\phi_j(\rr)d\rr}=&\frac{e^2}{4\kappa\sqrt{\pi}a^2_B}\exp\left(-\frac{1}{4 a^2_B}|\bm{R}_i-\bm{R}_j|^2-\frac{1}{8a^2_B}|\bm{R}_i+\bm{R}_j-2\bm{R}_C|^2\right)\\
&\times\text{I}_0\left(\frac{1}{8a^2_B}|\bm{R}_i+\bm{R}_j-2\bm{R}_C|^2\right)
\end{aligned}\label{eq:AppBimp}
\end{equation}
where $\text{I}_0$ is the zeroth-order modified Bessel function of the first kind. Note here that both wave functions $\phi_i(\rr)$ and $\phi_j(\rr)$ appearing in Eq.~\eqref{eq:AppBimp} are wave functions of quantum-dot electrons. The impurity only manifests itself as the additional Coulomb potential in the integrand.

\section{Charge noise caused by an impurity close to the quantum dot}\label{appc}

\begin{figure}
  \includegraphics[width=0.4\columnwidth]{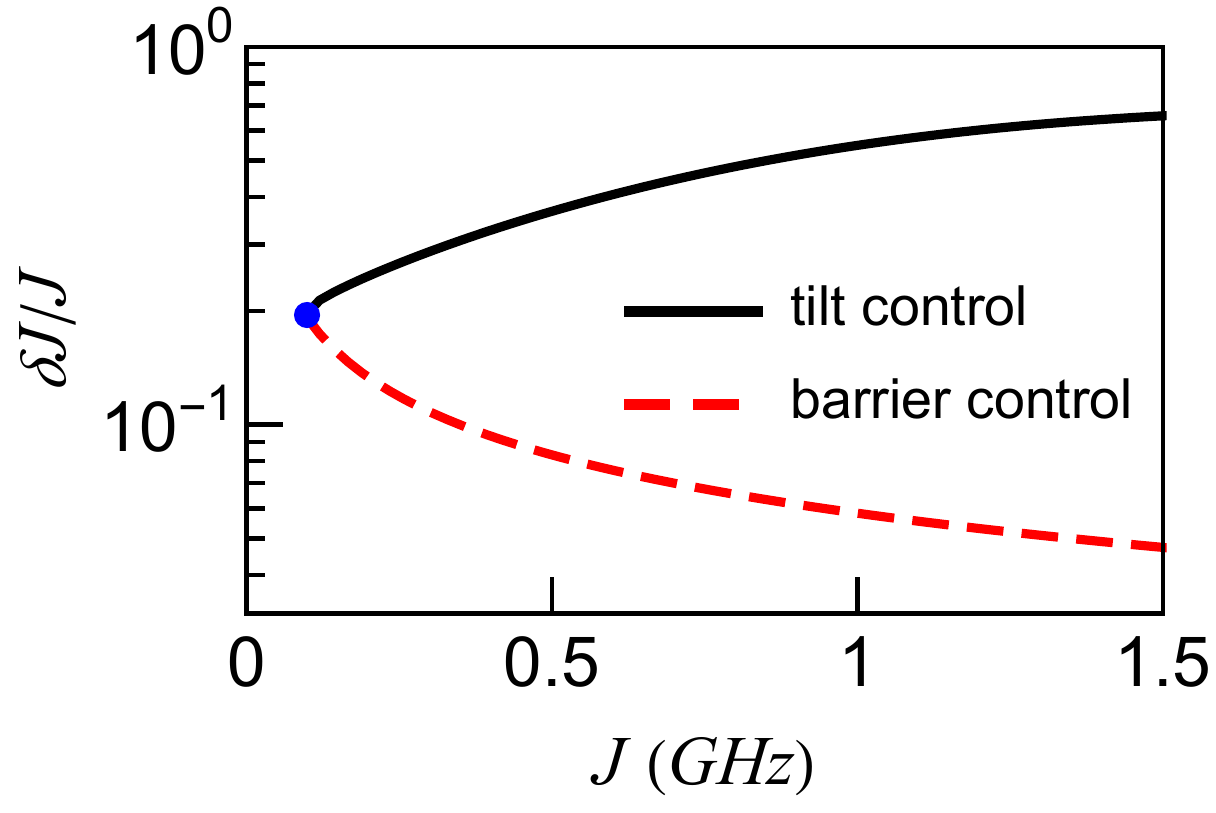}
  \caption{The relative charge noise $\delta J/J$ v.s.~$J$. Black solid line: the result for the tilt control case. Red (gray) dashed line: the result under barrier control. Comparison of the different exchange interaction $\delta J/J$ caused by impurity with tilt and barrier control. Note the log scale of the $y$-axis. The impurity is positioned at $\bm{R}_c= (-1.5a, 0.5a)$ and has charge $-0.01e$.  Parameters: $a$ = $100$ nm, $\hbar \omega_0$ = $100$ $\mu$eV.}
\label{fig:7}
\end{figure}

For all results shown in the main text, the impurity is considered to be at a reasonable distance away from the quantum dot. Nevertheless, our main conclusion still holds true even if the impurity is close to the quantum dot. Figure~\ref{fig:7} shows  the calculated relative charge noise caused by an impurity positioned at $\bm{R}_c= (-1.5a, 0.5a)$. Note that the charge of the impurity is $-0.01e$. Should an impurity with charge $-e$ be considered in this case, the noise it is causing under the tilt control would be much larger than the exchange interaction itself, an impractical situation. We therefore keep the impurity charge small when it is close to the quantum dot. It should be clear from Fig.~\ref{fig:7} that even if the impurity is close to the quantum dot, the relative charge noise will decrease with increasing $J$ when the barrier control is implemented, but will instead increase if tilt-controlled.

\twocolumngrid


%

\end{document}